\documentclass[manuscript]{aastex}

\shorttitle{{\em Kepler} Target Stars}
\shortauthors{Batalha et al.}

\begin{document}

\title{Selection, Prioritization, and Characteristics of {\em Kepler} Target Stars}


\author{Natalie M. Batalha}
\affil{Department of Physics and Astronomy, San Jose State University,
	San Jose, CA 95192}
\email{Natalie.Batalha@sjsu.edu}

\author{William J. Borucki, David G. Koch, Stephen T. Bryson, and Michael.R. Haas}
\affil{NASA Ames Research Center, Moffett Field, CA}

\author{Timothy M. Brown}
\affil{Las Cumbres Observatory Global Observatory Telescope Network,
	Goleta, CA 93117}

\author{Douglas A. Caldwell}
\affil{SETI Institute, Mountain View, CA 94043}

\author{Jennifer R. Hall}
\affil{Orbital Sciences Corporation/NASA Ames Research Center, Moffett Field, CA 94035}

\author{Ronald L. Gilliland}
\affil{STScI, Baltimore, MD 21218}

\author{David W. Latham and Soren Meibom}
\affil{Harvard-Smithsonian, CfA, Cambridge, MA 02138}

\and

\author{David G. Monet}
\affil{U.S. Naval Observatory, Flagstaff, AZ 86001}


\begin{abstract}
The {\em Kepler Mission} began its 3.5-year photometric monitoring campaign in May 2009 on a select group of approximately 150,000 stars.  The stars were chosen from the $\sim$half million in the field of view that are brighter than 16th magnitude.  The selection criteria are quantitative metrics designed to optimize the scientific yield of the mission with regards to the detection of Earth-size planets in the habitable zone.  This yields more than 90,000 G-type stars on or close to the Main Sequence, $>20,000$ of which are brighter than 14th magnitude. At the temperature extremes, the sample includes approximately 3,000 M-type dwarfs and a small sample of O and B-type MS stars ($<200$).  Small numbers of giants are included in the sample which contains $\sim$5,000 stars with surface gravities $\log(g) < 3.5$.  We present a brief summary of the selection process and the stellar populations it yields in terms of surface gravity, effective temperature, and apparent magnitude.  In addition to the primary, statistically-derived target set, several ancillary target lists were manually generated to enhance the science of the mission, examples being: known eclipsing binaries, open cluster members, and high proper-motion stars.

\end{abstract}


\keywords{stars: fundamental parameters --- techniques: photometric --- catalogs}

\section{Introduction}
\label{sec:intro}

{\em Kepler's} primary objective is to detect Earth-size planets in the habitable zone of Main Sequence stars.  The large field of view (FOV) targeting the rich star field in Cygnus ($\sim 10$ degrees off the galactic plane) was carefully chosen to optimize the science yield with regards to that specific objective \citep{fovBatalha,fovJenkins}. Even so, approximately 40\% of the stars brighter than R=16.5 that fall on silicon are expected to be giants and supergiants according to simulations \citep{besancon} of the stellar populations in that region of the sky \citep{fovBatalha}.  During the mission design phase, it was recognized that significant improvements in efficiency could be realized by pre-selecting the target stars.  On the order of $10^5$ stars brighter than 16th magnitude in the {\em Kepler} passband (denoted herein as $Kp$) were expected to have the right properties for detectability of terrestrial-size planets, implying that less than $10\%$ of the pixels need to be downloaded from the spacecraft.  Furthermore, pre-selection of the Main Sequence stars eliminates entire classes of astrophysical false-positives, reducing the resources required by the ground-based follow-up teams.  

Beginning with the pre-launch stellar classification effort (Section~\ref{sec:kic}), we describe the methodology for cherry-picking 150,000 stars from the $\sim 0.5$ million in the field of view brighter than 16th magnitude (Section~\ref{sec:selection}). We summarize the criteria used to prioritize the stellar targets (Section~\ref{sec:priority}) and present a summary of the characteristics of the sample (Section~\ref{sec:char}).  The stellar sample described here is specific to the Quarter 4 (mid-December 2009 through mid-March 2010) targets.  The populations differ slightly from those observed May-December 2009 \citep{batalhaIAU} due to revised estimates of the expected per cadence noise and a change in the definition of the crowding metric.  The target selection described herein utilizes a crowding metric computed using a set of pixels that define the optimal photometric aperture \citep{prfBryson} for each individual star as opposed to a fixed $11\times11$ pixel aperture centered on each star. Lastly, we provide an overview of the ancillary target lists drawn from specialized sources (Section~\ref{sec:ancillary}). While these comprise a small percentage of the sample, they contribute significantly to the primary objectives of the mission.  

\section{The {\em Kepler Input Catalog}}
\label{sec:kic}
A ground-based observing campaign was initiated in 2004 to determine the apparent
magnitude ($Kp$), surface gravity, effective temperature, metallicity, and interstellar
extinction of stars in the {\em Kepler} field from broad and intermediate band photometry.  An estimate of the stellar radius and mass follows via isochrone interpolation.  The calibrated photometry and resulting stellar parameters are archived in the {\em Kepler Input Catalog} (KIC) which is publicly available at the Multi-Mission Archive at Space Telescope Science Institute \footnote{http://archive.mast.edu/kepler}.  We provide, here, a brief summary of the process leading to stellar classification.  A full description of the algorithms is available at \url{http://www.cfa.harvard.edu/kepler/kic/kicindex.html}.   

A custom photometer (KeplerCam) was built for the 48-inch telescope at Whipple Observatory on Mt. Hopkins utilizing a Fairchild CCD486 \citep{kepcam}.  The campaign required multiple visits to $>1,600$ pointings \citep{kicLatham} covering the {\em Kepler} FOV and the open cluster, M67 (used for color calibrations and reliability tests).  Multiple exposures were taken in the (Sloan-like) $g, r, i, z$ filters as well as a custom intermediate-band filter, D51, centered at 510 nm (Mg b line).  A small number of pointings included $u$ and $G_{red}$ (a custom intermediate-band filter centered at 432 nm) observations.  However, the long exposure times required for the requisite SNR were prohibitively costly, and these filters were dropped during the earlier stages of the program.  Surface gravity dependence is largely provided by the D51 measurements. For the M-type stars, $J-K$ is a more sensitive diagnostic.  The {\em 2MASS} J,H,K magnitudes are federated with the calibrated $g,r,i,z$, and D51 magnitudes for the purpose of stellar classification. Corrections for atmospheric extinction are based on nightly observations of $\sim$1500 secondary photometric standards within the Kepler field.  These in turn are tied to $>$300 stars from SDSS DR1 \citep{smith02}, selected for distribution on the sky and range in color.
  
The $g,r,i,z$, D51, and J,H,K magnitudes yield 7 independent color measurements that are calibrated by zero-point offsets derived from cluster observations and Sloan standard stars. These 7 observables are used together with the $r$ magnitude and the galactic latitude to derive 11 parameters, namely: effective temperature (T$_{eff}$), surface gravity ($\log(g)$), metallicity ($\log(Z)$), mass, radius, luminosity, bolometric correction, distance, interstellar extinction ($A_r$ and $A_V$), and reddening ($E_{B-V}$).  Functions relating these parameters (e.g. extinction laws \citep{cardelli}, the radius dependence on mass and surface gravity, the luminosity dependence on effective temperature and radius, and the apparent magnitude dependence on luminosity, the bolometric correction, interstellar extinction, and distance) reduce the number of unknowns from 11 to 5.

Stellar classification requires choosing among model stellar atmospheres that best fit the observables.  The \citet{kurucz} models provide flux as a function of wavelength for temperatures ranging from 3500 to 50000 K, surface gravities of 0 to 5.5 dex, and metallicities of $-3.5$ to $+0.5$. Fluxes are transformed to magnitudes by multiplying by an estimate of the CCD response and filter transmission and integrating over wavelength. The model colors (formed taking relevant magnitude differences) are calibrated to the cluster observations using a zero-point offset and a term linear in ($g-r$). The evolutionary tracks of \citet{girardi00} are used to constrain stellar luminosity (and, ultimately, radius and mass) given effective temperature and surface gravity (assuming solar metallicity).

The photometric data are minimally sufficient for obtaining reliable surface gravity determinations.  The parameter estimation is, therefore, carried out using additional astrophysical information to form Bayesian priors.  A star's physical parameters (given as the vector $\vec{x}$) are those which maximize the posterior probability of obtaining the observed magnitudes and colors (given as the vector $\vec{q}$).  The posterior probability, $P(\vec{x}|\vec{q})$, is given by:
\begin{equation}
P(\vec{x}|\vec{q})=[P_0(x)P(\vec{q}|\vec{x})]/P_0(\vec{q}) .
\end{equation}
The denominator in this expression, the a priori probability of observing photometric indices given by $\vec{q}$, is a normalization that does not depend upon the parameters $\vec{x}$ of interest; for purposes of maximizing the posterior probability, it may be ignored. 

Assuming independent, Gaussian-distributed errors, one can write
\begin{equation}
P(\vec{q}|\vec{x})=exp(-\chi^2) ,
\end{equation}
where $\chi^2$ is the usual goodness-of-fit statistic.  The prior probability, $P_0(\vec{x})$, is parametrized as
\begin{equation}
P_0(\vec{x})=P_{HR}[T_{eff},\log(g)]P_Z[\log{Z}]P_z(z) ,
\end{equation}
where $P_{HR}$ is the probability of finding a star in a given region of the HR diagram (parametrized in terms of effective temperature and surface gravity), $P_Z$ is the probability of having a metallicity $Z$, and $P_z$ is the probability of being situated at a height, $z$, above the galactic plane.  $P_{HR}$ is constructed from the 9,590 Hipparcos stars having parallaxes known to better than 10\%.  The metallicity distribution is taken from the compilation of \citet{nordstrom} for 14,000 bright, nearby stars.  The stellar density is assumed to decrease exponentially with a scale height of 300 pc.  It allows for the cubic increase in volume with increasing distance.

The advantage of employing the Bayesian method is that implausible stellar classifications are ruled out a priori.  The disadvantage is that stars with rare properties are liekly to be misclassified.  For our primary objective -- that of distinguishing giants from dwarfs -- the Bayesian approach can be expected to work well.  Stellar parameters are derived for $\sim$80\% of the stars brighter than 17th magnitude.  The classification can fail for a multitude of reasons, the most common of which is related to the completeness of the photometric measurements.  Classification is not attempted if there are fewer than 2 valid optical magnitudes or 3 NIR magnitudes in the database.  This completeness rate is independent of magnitude for stars brighter than $r=17$.

The photometry and derived parameters are archived in the KIC together with results from other photometric survey catalogs (e.g. 2MASS, USNO-B, Tycho, and Hipparcos) in order to provide a full census of point-sources in the field down to SDSS $g=20$ corresponding to the completeness of the USNO-B catalog \cite{monet03}.  The stellar parameters form the basis of target selection and prioritization.

During the commissioning phase of {\em Kepler} operations, 9.7 days of 30-minute cadence photometry was collected on a sample of 52,496 of the brightest stars (excluding those known to be highly crowded by neighboring stars). From this sample, one-thousand bright, unsaturated, known giants (well-studied astrometric grid stars) and MS dwarfs (as indicated by KIC classifications and, in some cases, confirmed by Hipparcos parallaxes) were used to assess the reliability of the classifications in the {\em Kepler Input Catalog}.  A study of 7 days of HST photometry (5 samples per 96-minute orbit) of bulge stars suggests that high-precision light curves at sufficiently high cadence can be used as a luminosity class discriminator \citep{gilliland08}. The RMS variability of the giant light curves is typically at least one order of magnitude larger than that of dwarfs.  Giants also present distinct frequency characteristics: quasi-periodic variations with several distinct frequencies over a narrow range while dwarfs exhibit predominantly white noise over a 10-day window.  This photometric luminosity class discriminator was applied to the {\em Kepler} light curves of the giant/dwarf control samples.  The results were inspected manually to assess the reliability of the stellar classifications in the {\em Kepler Input Catalog} with regards to giant/dwarf separation.  The KIC designations agree with the variability discriminator $> 90\%$ of the time for unsaturated stars brighter than 13th magnitude.

\section{Target Selection}
\label{sec:selection}

We approach target selection by computing, for every KIC-classified star in the field, the radius of the smallest planet detectable at the 7.1-sigma level (threshold at which we expect less than one statistical false-positive) over the mission lifetime.  The total SNR is the ratio of the transit depth (which depends on the ratio of the planet to star surface area) to the total noise, $\sigma_{tot}$, in the folded and binned light curve.  The calculation requires knowledge of the stellar parameters  as well as the instrument performance. The per-cadence noise is modeled and scaled by $1/\sqrt{N_{tr}}$ where $N_{tr}$ is the number of transits observed over the 3.5-year mission duration and by $1/\sqrt{N_{sample}}$ where $N_{sample}$ is the number of 30-minute samples per transit event to give $\sigma_{tot}$.  The number of transits depends on the orbital period for a given stellar mass, $M_*$, and semi-major axis, $a$, while the number of samples per transit depends on the transit duration ($t_{tr}$) which, assuming a circular orbit, is given by:
\begin{equation}
t_{tr}=\frac{\pi}{4}t_0 = 2R_*\sqrt{\frac{a}{GM_*}}
\end{equation}
where $t_0$ is the central transit duration.  It is scaled by $\pi/4$ to give the mean transit duration for stars randomly distributed in inclination.

The minimum detectable planet radius, $R_{p,min}$, is computed at three semi-major axes: 1) the inner radius of the habitable zone (HZ), 2) half that distance, and 3) five stellar radii.  A planet is not considered detectable unless the total SNR exceeds 7.1-sigma and at least three transits occur over the 3.5-year mission lifetime.  We use the standard solar-system inner HZ radius of \citet{hz} (0.95 AU) and scale it by the ratio of the stellar luminosity (computed from the $T_{eff}$ and $R_*$ archived in the KIC) to that of the Sun. 

The completeness of the KIC with regards to point sources down to SDSS $g=20$ allows us to simulate the sky and estimate the crowdedness of every point source.  Crowding can lead to flux contamination in the photometric aperture of the target star.  The excess flux dilutes the transits in the photometric light curve, thereby reducing their detectability.  A crowding metric, $r$, is defined to be the flux of the target star, $F_*$, in the photometric aperture expressed as a fraction of the total flux (star plus background flux, $F_{bg}$, which includes contributions from zodiacal emission as well as background stars):
\begin{equation}
r = \frac{F_*}{F_* + F_{bg}} .
\end{equation}
The photometric aperture is defined as the set of pixels that optimizes the total SNR and is referred to as the {\em Optimal Aperture}.  It is dependent on the local Pixel Response Function (PRF), measured on-orbit during the commissioning period \citep{prfBryson}, as well as the distribution of stellar flux on the sky near the target.  The latter relies on information from the {\em KIC}.  An optimal aperture and crowding metric is computed for every potential target star.

\begin{figure}[h!]
\resizebox{\hsize}{!}{\includegraphics{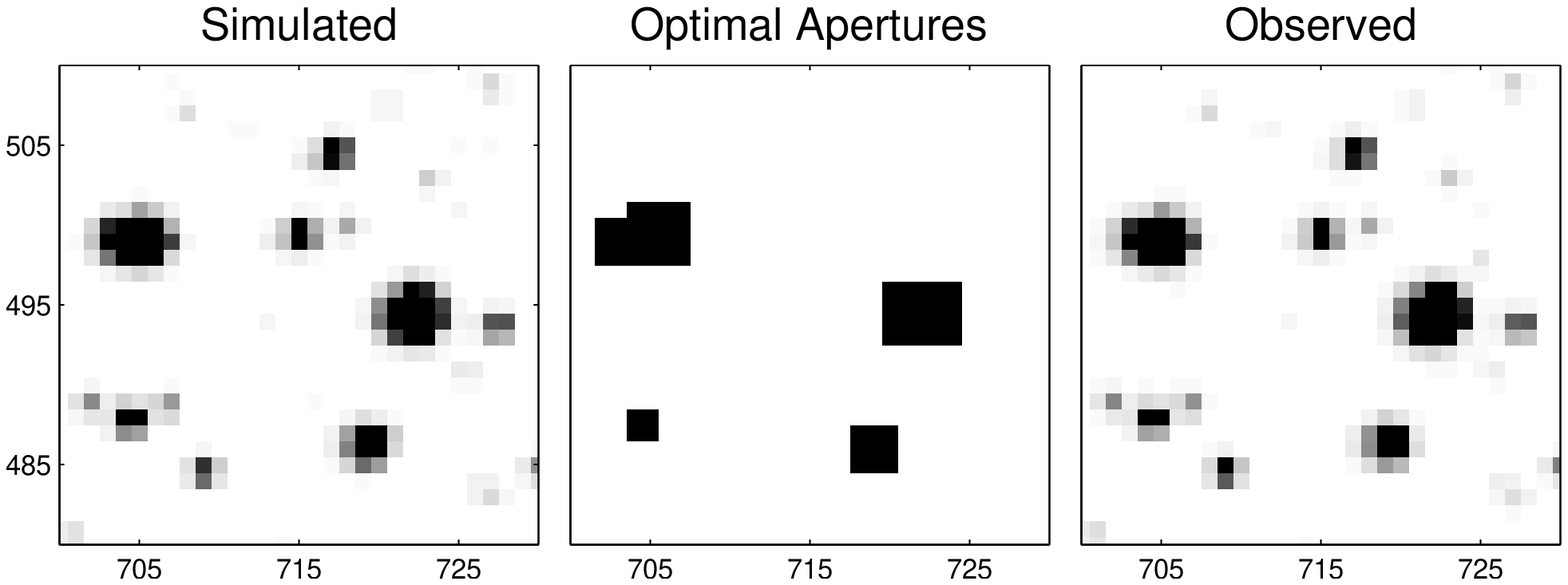}}
\caption{A comparison of the thumbnails of 1) a simulated image (left) and 2) a full-frame image taken during flight (right) shows that the sky is well-reproduced by models at even the pixel level.  Simulated images use the {\em Kepler Input Catalog} as input as well as models of the instrument characteristics produced during the commissioning period (e.g. Pixel Response Function, Focal Planet Geometry, etc).  The middle panel shows the mask that defines the optimal aperture pixels for the target stars in this region of the sky.}
\label{fig:ffiCoa}
\end{figure}

Figure~\ref{fig:ffiCoa} shows a small region ($2\times2$ arcmin) of a simulated image (left), sized to highlight individual stars. Also shown (right) is the same region of an actual full-frame image taken on orbit by {\em Kepler}.  Individual stars are well-reproduced by simulations as are the pixel-to-pixel brightness differences.  Note that the same gray-scale mapping and range is used for each image. The middle panel shows the same region with the optimal aperture pixels blackened out.  The diluted fractional transit depth, $f$, is related to the intrinsic fractional depth, $f_0$, via the crowding metric, $r$:
\begin{equation}
f=r \times f_0 .
\end{equation}

Combining the above formulation, the minimum detectable planet radius, $R_{p,min}$, is given by:
\begin{equation}
R_{p,min} = R_* \sqrt{\frac{7.1 \sigma_{tot}}{r}} ,
\end{equation}
where $R_*$ is the stellar radius.  More than 200,000 stars in {\em Kepler's} FOV have a minimum detectable planet radius smaller than two earth-radii ($2R_e$) at a semi-major axis corresponding to the inner HZ for that star as well as 3 or more transits over the lifetime of the mission.  However, more than $60\%$ are fainter than 15th magnitude and are not, consequently, amenable to high-precision RV follow-up. Clearly, a strict sorting on $R_{p,min}$ does not yield the desired population. A more suitable prioritization scheme is required.  

\section{Prioritization}
\label{sec:priority}

Table~\ref{priority:tb} lists the prioritization criteria, ordered from high to low priority, and the resulting cumulative star counts.
The highest priority targets are those for which a terrestrial-size planet is detectable in the HZ and for which 1-3 m/s radial velocity precision can be readily achieved with existing instruments and methods ($Kp \leq 14$). It is recognized that the very late-type stars will tend to be faint, especially in the optical, and that these populations will benefit considerably from the high-resolution infrared spectrometers and RV templates that are being developed to make use of the proportionately higher photon fluxes these objects emit in the IR.  Consequently, the next highest priority targets are those for which Earth-size planets are detectable in the HZ even if they are as faint as 15-16 magnitudes.  The fainter K and M dwarfs will be captured here. The HZ constraint is relaxed next, bringing in stars that did not meet the 3-transit criterion at $a=HZ$ as well as those that benefit from the additional SNR built up by observing more transits at the shorter period orbits.

The end product is a sample of $261,636$ stars brighter than 16th magnitude that fall on silicon for which a planet of radius $\leq 2 R_e$ is detectable (at acceptable confidence levels) over the duration of the mission.  We do not make an attempt to include predictions with regards to stellar variability in the noise models used for target selection (since so little is known about stellar variability across the HR diagram at this precision). Such a metric can be included in the future as we learn more about the target stars and will be important in the event we are forced to downselect.  The final target list is drawn from this prioritized list of stars, trimmed down to meet the mission constraints: number of targets $<170,000$ and number of pixels $<5.44$ million.

\section{Characteristics of the Target Stars}
\label{sec:char}

Table~\ref{tab:ms} reports the star counts from amongst the 150,000 highest priority targets, binned by apparent magnitude and effective temperature and separated by surface gravity ($\log g \geq 3.5$ and $\log g < 3.5$).  Some giants make it into the target sample when the combination of apparent magnitude and stellar radius allows for the detection of planets as small as $2R_e$ in an orbit as close as $5R_*$. The sample is dominated by G-type stars on or near the Main Sequence and stars fainter than 14th magnitude.  At the temperature extremes, we have $\sim3,000$ M-type Main Sequence stars and $< 200$ O and B-type stars.  OB-type stars are rare because of their short MS lifetimes and the fact that the field of view was chosen to avoid young stellar populations (OB associations and star forming regions).

Table~\ref{tab:remaining} reports star counts for the populations that did not satisfy the detectability criteria. These are the stars brighter than 16th magnitude which can be included in proposals for the Guest Observer Program \footnote{http://keplergo.arc.nasa.gov}. The targets brighter than 14th magnitude are almost exclusively evolved stars. Main Sequence stars not being observed by the mission are predominantly G-type stars fainter than 15th magnitude.

\section{Ancillary Target Lists}
\label{sec:ancillary}

Table~\ref{tab:ms} represents the statistically derived {\em PLANETARY} target list.  Several smaller, ancillary target lists have been constructed to ensure no high priority targets are missed. All known eclipsing binaries ({\em EB}) in the field ($> 600$) are included on an {\em EB} target list drawn from ground-based surveys \citep{asas,vulcan,hatnet} and from EB's already flagged in the SIMBAD database.  EB light curves will be combed for transit events and subjected to eclipse timing using techniques developed by {\em Kepler Participating Scientists}. There are four open clusters in the {\em Kepler} field of view: NGC 6866 ($\sim$0.5 Gyr), NGC 6811 ($\sim$1 Gyr), NGC 6819 ($\sim$2.5 Gyr), and NGC 6791 ($\sim$ 10 Gyr).  All known members are included in a {\em CLUSTER} target list.  The clusters will be used to extend the age-rotation relation beyond the age of the Hyades \citep{meibom}. This calibration will help to define the ages of stars with planets. The closest Main Sequence stars are explicitly added to the target list by including a) 141 high proper-motion K and M dwarfs from the LSPM-North catalog \citep{lepine}, b) stars with Hipparcos parallaxes that confirm their Main Sequence status ($\sim 200$ stars), and c) M-dwarfs from the Gliese catalog ($\sim 10$). Many of the targets captured in the above lists are also identified by the statistical selection process based on detectability metrics.  However, explicit inclusion ensures that these high priority targets will not be inadvertently dropped in future down-selects.

Finally, as noted in Section~\ref{sec:kic}, approximately $20\%$ of the stars in the {\em Kepler Input Catalog} brighter than $Kp=17$ lack effective temperature, surface gravity, and radius estimates.  Consequently, we began science operations observing $\sim 10,000$ unclassified stars brighter than 14th magnitude. The number has been reduced by applying the photometric luminosity class discriminator described in Section~\ref{sec:kic} to the data collected during commissioning and the first 33 days of science operations.  1,996 of the $\sim 10,000$ stars have been explicitly included as a result of the dwarf/giant discriminator while 2,272 of the $\sim 10,000$ stars have been explicitly excluded.  The remaining $\sim 6,000$ remain on the unclassified target list pending additional data. The unclassified stars that were determined to be giants comprise part of the Quarter 2 Dropped Target List.  The light curves of dropped targets are made available to the public 60 days after release \citep{haas}.  

\section{Summary}
\label{sec:summary}

{\em Kepler} is observing more than 150,000 stars during the first year of mission operations, more than 90,000 of which are G-type stars on or near the Main Sequence.  10,575 are G-type stars brighter than 13th magnitude. 28,519 are brighter than 14th magnitude. Over $90\%$ of the target stars are selected based on detectability metrics that suggest a terrestrial-size planet ($R < 2R_{e}$) is detectable in 3.5 years.  The detectability metrics are derived from stellar properties in the {\em Kepler Input Catalog}.  Stellar classifications (surface gravity, effective temperature, and the inferred stellar radius) are complete at the $80\%$ level, independent of magnitude.  A photometric luminosity class discriminator is applied to 1,000 Main Sequence and 1,000 Giant light curves (as indicated by KIC classifications and confirmed, in some cases, by Hipparcos parallaxes) from commissioning data.  We report agreement between the KIC and photometric discriminators $>90\%$ of the time for stars brighter than 13th magnitude. Priority on the target list is given to stars that can be followed up with high-precision radial velocity measurements from current ground-based facilities. High priority is also assigned to the fainter ($14 \leq Kp < 16$) K and M-type dwarfs that will benefit from future IR spectrometers.  Ancillary target lists are generated for special purposes (e.g. to detect planets around EBs, to extend the age-rotation relation for Main Sequence stars, and to ensure that the closest Main Sequence stars are included).  Finally the characteristics of the stars that are not being observed are provided for prospective Guest Observers. 

\acknowledgments

Funding for this Discovery mission is provided by NASA's Science Mission Directorate.





\newpage

\newpage



\begin{table}\def~{\hphantom{0}}
\begin{center}
\caption{Star counts as a function of effective temperature and apparent magnitude ($Kp$) for the 150,000 highest priority target stars.}
\label{tab:ms}
\begin{tabular}{lccccccccr}\hline
MAG &  10500 & 9500 & 8500 & 7500 &  6500 &  5500 &  4500 &  3500 &    TOTAL \\
\hline
\hline
\multicolumn{10}{c}{$\log g \geq 3.5$} \\
\hline
 6.50 &     1 &     0 &     1 &     2 &     0 &     1 &     0 &     0 &        5 \\
 7.50 &     1 &     8 &     9 &     6 &     8 &     6 &     0 &     0 &       38 \\
 8.50 &     8 &    20 &    25 &    24 &    49 &    15 &     7 &     8 &      156 \\
 9.50 &     9 &    31 &    81 &    66 &   116 &    88 &    11 &     4 &      406 \\
10.50 &    27 &    37 &   100 &   209 &   405 &   359 &    40 &     9 &     1186 \\
11.50 &    24 &    58 &   171 &   398 &  1499 &  1356 &   158 &    37 &     3701 \\
12.50 &    30 &    44 &   231 &   676 &  4146 &  4760 &   626 &    62 &    10575 \\
13.50 &    34 &    53 &   170 &   747 &  9279 & 15866 &  2213 &   157 &    28519 \\
14.50 &     3 &     0 &     0 &     0 &  4855 & 29352 &  4227 &   554 &    38991 \\
15.50 &     7 &     4 &     0 &     0 &  4449 & 42627 & 12093 &  1961 &    61141 \\
TOTAL &   144 &   255 &   788 &  2128 & 24806 & 94430 & 19375 &  2792 &   144718 \\
\hline
\multicolumn{10}{c}{$\log g < 3.5$} \\
\hline
 6.50 &     0 &     0 &     0 &     0 &     0 &     0 &     0 &     0 &        0 \\
 7.50 &     0 &     0 &     1 &     1 &     2 &     2 &     7 &     0 &       13 \\
 8.50 &     0 &     0 &     1 &     2 &     2 &     9 &    80 &     2 &       96 \\
 9.50 &     0 &     0 &     2 &    15 &     2 &    27 &   220 &     1 &      267 \\
10.50 &     1 &     0 &     5 &    21 &     7 &    99 &   452 &     2 &      587 \\
11.50 &     0 &     0 &     1 &    25 &    11 &   186 &   674 &     2 &      899 \\
12.50 &     0 &     0 &     1 &    12 &    14 &   347 &  1114 &     0 &     1488 \\
13.50 &     0 &     0 &     0 &     6 &     5 &   518 &  1403 &     0 &     1932 \\
14.50 &     0 &     0 &     0 &     0 &     0 &     0 &     0 &     0 &        0 \\
15.50 &     0 &     0 &     0 &     0 &     0 &     0 &     0 &     0 &        0 \\
TOTAL &     1 &     0 &    11 &    82 &    43 &  1188 &  3950 &     7 &     5282 \\
\hline
\end{tabular}
\end{center}
\end{table}

\begin{table}\def~{\hphantom{0}}
\begin{center}
\caption{Star counts as a function of effective temperature and apparent magnitude ($Kp$) for the remaining KIC stars with stellar classifications that are brighter than 16th magnitude.}
\label{tab:remaining}
\begin{tabular}{lccccccccr}\hline
MAG &  10500 & 9500 & 8500 & 7500 &  6500 &  5500 &  4500 &  3500 &    TOTAL \\
\hline
\hline
\multicolumn{10}{c}{$\log g \geq 3.5$} \\
\hline
 6.50 &     0 &     0 &     0 &     0 &     0 &     0 &     0 &     0 &        0 \\
 7.50 &     0 &     0 &     0 &     0 &     0 &     0 &     0 &     0 &        0 \\
 8.50 &     0 &     0 &     0 &     0 &     0 &     0 &     0 &     0 &        0 \\
 9.50 &     0 &     0 &     0 &     0 &     0 &     0 &     0 &     0 &        0 \\
10.50 &     0 &     0 &     0 &     0 &     0 &     0 &     0 &     0 &        0 \\
11.50 &     0 &     0 &     0 &     0 &     0 &     0 &     0 &     0 &        0 \\
12.50 &     0 &     0 &     0 &     0 &     0 &     0 &     0 &     0 &        0 \\
13.50 &     0 &     0 &     0 &     0 &     0 &     0 &     0 &     0 &        0 \\
14.50 &    50 &    74 &   212 &   707 & 11340 & 14083 &  2998 &     1 &    29465 \\
15.50 &    66 &    87 &   277 &  1038 & 25476 & 55506 &  7737 &   107 &    90294 \\
TOTAL &   116 &   161 &   489 &  1745 & 36816 & 69589 & 10735 &   108 &   119759 \\
\hline
\multicolumn{10}{c}{$\log g < 3.5$} \\
\hline
 6.50 &     0 &     0 &     0 &     0 &     0 &     0 &     1 &     0 &        1 \\
 7.50 &     0 &     0 &     0 &     0 &     0 &     0 &     4 &     2 &        6 \\
 8.50 &     0 &     0 &     0 &     0 &     0 &     1 &    81 &    52 &      134 \\
 9.50 &     0 &     0 &     0 &     0 &     0 &     8 &   295 &   125 &      428 \\
10.50 &     0 &     0 &     0 &     0 &     0 &    37 &   989 &   273 &     1299 \\
11.50 &     0 &     0 &     0 &     0 &     1 &    98 &  2691 &   404 &     3194 \\
12.50 &     0 &     0 &     0 &     1 &     1 &   209 &  5673 &   505 &     6389 \\
13.50 &     0 &     0 &     1 &     1 &     2 &   375 & 10254 &   424 &    11057 \\
14.50 &     0 &     0 &     0 &     2 &    13 &  1251 & 15328 &   311 &    16905 \\
15.50 &     0 &     0 &     0 &     5 &    11 &  1672 & 15657 &   252 &    17597 \\
TOTAL &     0 &     0 &     1 &     9 &    28 &  3651 & 50973 &  2348 &    57010 \\
\hline
\end{tabular}
\end{center}
\end{table}

\begin{table}\def~{\hphantom{0}}
   \begin{center}
   \caption{Prioritization: cumulative star counts}
   \label{tab:priority}
   \begin{tabular}{lr}\hline
Criteria & Counts \\
\hline
$N_{tr} \geq 3; R_{p,min} \leq 1 R_e; a=HZ; Kp < 13$~~~~~ & 3742 \\
$N_{tr} \geq 3; R_{p,min} \leq 2 R_e; a=HZ; Kp < 13$ & 4169 \\
$N_{tr} \geq 3; R_{p,min} \leq 1 R_e; a=HZ; Kp < 14$ & 9323 \\
$N_{tr} \geq 3; R_{p,min} \leq 2 R_e; a=HZ; Kp < 14$ & 17375 \\
$N_{tr} \geq 3; R_{p,min} \leq 1 R_e; a=HZ; Kp < 15$ & 22714 \\
$N_{tr} \geq 3; R_{p,min} \leq 1 R_e; a=HZ; Kp < 16$ & 25610 \\
$N_{tr} \geq 3; R_{p,min} \leq 1 R_e; a=\frac{1}{2} HZ; Kp < 14$ & 30802 \\
$N_{tr} \geq 3; R_{p,min} \leq 2 R_e; a=\frac{1}{2} HZ; Kp < 14$ & 46366 \\
$N_{tr} \geq 3; R_{p,min} \leq 2 R_e; a=5R_*; Kp < 14$ & 58103 \\
$N_{tr} \geq 3; R_{p,min} \leq 2 R_e; a=HZ; Kp < 15$ & 91755 \\
$N_{tr} \geq 3; R_{p,min} \leq 2 R_e; a=HZ; Kp < 16$ & 154008 \\
$N_{tr} \geq 3; R_{p,min} \leq 2 R_e; a=5R_*; Kp < 15$ & 183563 \\
$N_{tr} \geq 3; R_{p,min} \leq 2 R_e; a=5R_*; Kp < 16$ & 261636 \\
   \end{tabular}
   \end{center}
\end{table}

\end{document}